\documentclass[eqsecnum,twocolumn,aps,
showpacs,amsmath,amssymb,floatfix,
longbibliography,superscriptaddress]{revtex4-1}
\usepackage{graphicx}
\DeclareGraphicsExtensions{.pdf}
\usepackage{amsmath,amssymb,bbold,bm,color}
\usepackage{float}
\usepackage{epstopdf}
\usepackage[dvipsnames]{xcolor}

\newcommand{\br}{{\bm r}}

\newcommand{\bdel}{{\bm \delta}}

\newcommand{\cT}{{\cal T}}

\newcommand{\cH}{{\cal H}}

\newcommand{\bee}{\begin{equation}}
\newcommand{\ee}{\end{equation}}

\def\sgn{\mathop{\rm sgn}\nolimits}

\begin{document}

\title{Mimicking black hole event horizons in atomic and solid-state systems}

\author{M. Franz}
\affiliation{Department of Physics and Astronomy, University of
British Columbia, Vancouver, BC, Canada V6T 1Z1}
\affiliation{Quantum Matter Institute, University of British Columbia,
  Vancouver BC, Canada V6T 1Z4}
\author{M. Rozali}
\affiliation{Department of Physics and Astronomy, University of
British Columbia, Vancouver, BC, Canada V6T 1Z1}

\date{\today}

\begin{abstract} 

Holographic quantum matter exhibits an intriguing connection between
quantum black holes and more conventional (albeit strongly interacting) quantum many-body systems. This connection is manifested in the study of their thermodynamics,
statistical mechanics and many-body quantum chaos.  After explaining some of those connections and their significance, we focus on the most promising example to date of holographic quantum matter, the
family of Sachdev-Ye-Kitaev (SYK) models. Those are simple
quantum mechanical models that are thought to realize,
holographically, quantum black holes. We review and assess various proposals for experimental realizations of the SYK models. 
 Such experimental realization offers the exciting prospect of
accessing black hole physics, and thus addressing many mysterious questions in quantum gravity, in tabletop experiments.

\end{abstract}

\date{\today}

\maketitle

\section{Introduction}

Among the greatest challenges facing modern physical sciences is the
relation between quantum mechanics and gravitational physics, described classically by Einstein's general theory of
relativity. Well known paradoxes arise when the two theories are
combined, for example when describing event horizons of black holes
\cite{Hawking1976}. Indeed, those paradoxes arise even though both general relativity
and quantum mechanics should be well within their regime of validity
\cite{Almheiri2013}. 

Perhaps the most promising avenue to resolving these paradoxes
is rooted in the holographic principle \cite{Susskind1995,tHooft2001}, which posits that quantum
gravity degrees of freedom in a $(d + 1)$-dimensional spacetime “bulk”
can be represented by a non-gravitational many-body system defined on its $d$-dimensional
boundary. By now there are many well-understood examples of such ``holographic dualities"  providing concrete realizations of that principle. Thus, using such holographic dualities, mysterious questions pertaining to gravitational physics in the quantum regime can be addressed in a context which is much better understood, that of quantum many-body physics.

Part of the difficulty in quantum gravity research is the dearth of relevant experiments. For example, experimental tests of the quantum gravity ideas using
astrophysical black holes are impractical, as the relevant energy scale is inaccessible. 
Instead one can turn to “holographic quantum
matter” \cite{Hartnoll2016} -- this is a class  of strongly interacting systems
that occur in certain solid-state or atomic systems, in which interactions between individual constituent
particles are so strong that the conventional quasi-particle description fails completely.  Rather, these systems exhibit properties of holography, in that they behave
effectively as black hole horizons in a higher dimensional quantum gravitational theory.  This opens up the exciting possibility of testing quantum gravity ideas experimentally, as holographic quantum matter is more amenable to experimental study, at least in principle.

This review focuses on recent developments aimed at theoretically understanding and realizing in the
laboratory a specific class of systems that give rise to holographic
quantum matter. They are described by the so called Sachdev-Ye-Kitaev
(SYK) models \cite{SY1996,Kitaev2015,Maldacena2016} which are ``dual", in the sense described above, to
a 1+1 dimensional  gravitating "bulk" containing a black hole.  We first review some of the background
material on black hole thermodynamics, statistical mechanics and its
relation to many-body quantum chaos. We then introduce the SYK model
and discuss some of its important physical properties that furnish
connections to holography and black holes. Finally we review proposals
for the experimental realization of this model using atomic and solid
state systems and assess critically their prospects and feasibility. We end by describing the outlook and potential significance of this realization of holographic dualities.

\section{Black Hole Materials}

The relation of black hole physics, the hallmark of quantum gravity research, to materials science is extremely  surprising. The story begins in early studies of quantum gravity in the 1970s. Famously there is some tension between fundamental concepts of General Relativity and Quantum Mechanics, and that tension manifest itself most clearly when black holes of General Relativity are treated as quantum mechanical objects. In classical gravity, black holes are characterized  by having a surface of no return, the so-called event horizon.  Any information crossing that surface and falling into the black hole is lost to observers who stay safely away from the black hole. However, that defining feature of black hole is modified quantum mechanically. Indeed, the celebrated work of Hawking \cite{Hawking:1974sw} established that quantum mechanically black holes are no longer black, they glow with what came to be called the Hawking radiation.

\subsection{Black Hole Thermodynamics}

The discovery of Hawking radiation finalized a conceptual revolution
of our understanding of black hole quantum mechanics. Hawking
radiation is thermal, like that of a black body, so black holes have a
temperature. Earlier Jacob Bekenstein \cite{Bekenstein:1973ur} has
speculated that the surface area of a black hole behaves analogously
to an entropy (for example it always increases); apparently this was
not merely an analogy.  Surprisingly black holes, as viewed by an
outside observer, behave exactly as thermodynamic systems
\cite{Bardeen:1973gs}. They have static properties: energy, charge,
entropy, temperature. They have dynamical properties such as viscosity
and conductivity. An outside observer who does experiments on black holes views them as chunks of materials with collective properties which obey all the usual relations of thermodynamics.

Indeed, there is a wide variety of possibly black holes in various gravitational theories, in various spacetime dimensions and with diverse matter content,  and correspondingly a wide variety of black hole ``materials". To explain that diversity, one had to wait for a microscopic theory of quantum black holes. 

\subsection{Black Hole Statistical Mechanics}

In the mid-1990s string theory, a leading candidate for a theory of quantum gravity, has reached the level of development needed to provide a microscopic statistical mechanics description of a large class of black holes \cite{Strominger:1996sh}.  
Within string theory, gravitational interactions emerge naturally within an inherently quantum mechanical framework. It became possible then to discuss, in great quantitative detail, static and dynamical properties of quantum black holes. The flurry of activity in the following decade established a description of quantum black holes which is, perhaps surprisingly, fairly conventional. Within this framework black holes are described as quantum many-body systems, composed of microscopic constituents interacting strongly, to form a quantum liquid whose thermodynamics forms the gravitational description of the system. In this sense gravity arises from coarse-graining, exactly like a thermodynamic description of conventional materials.

The microscopic description makes it clear that black holes obey the rules of quantum mechanics, in particular they avoid Hawking's information loss paradox \cite{Hawking1976}. Nevertheless, it is still unclear {\it how} the paradox is avoided, in other words how resolution of the paradox can be  phrased in a purely gravitational language. Recently this set of questions was sharpened and cast in  the language of quantum information processing \cite{Almheiri2013}. Understanding black holes as quantum computers is an exciting current direction in quantum gravity, which is expected to provide valuable clues on how a gravitational description emerges for black holes, while avoiding information loss and related puzzles. 

\subsection{Box 1: Gauge/Gravity Duality}

The microscopic description of black holes is at its most developed
stage in the context of the AdS/CFT correspondence, also known as the
gauge-gravity or holographic duality. Here AdS refers to anti-de Sitter space (see
below) and CFT denotes conformal field theory. In this set of examples, the black hole is immersed in a gravitational potential (arising due to a negative cosmological constant), which acts to regulate some problematic long distance behaviour unrelated to the questions we are interested in. When thus regulated, it turns out the microscopic description is in terms of well-known many-body concepts, namely gauge theories of the form that constitutes the standard model of particle physics, as well as many other interesting systems appearing in the condensed matter context.
\begin{figure}[t]
\includegraphics[width = 7.6cm]{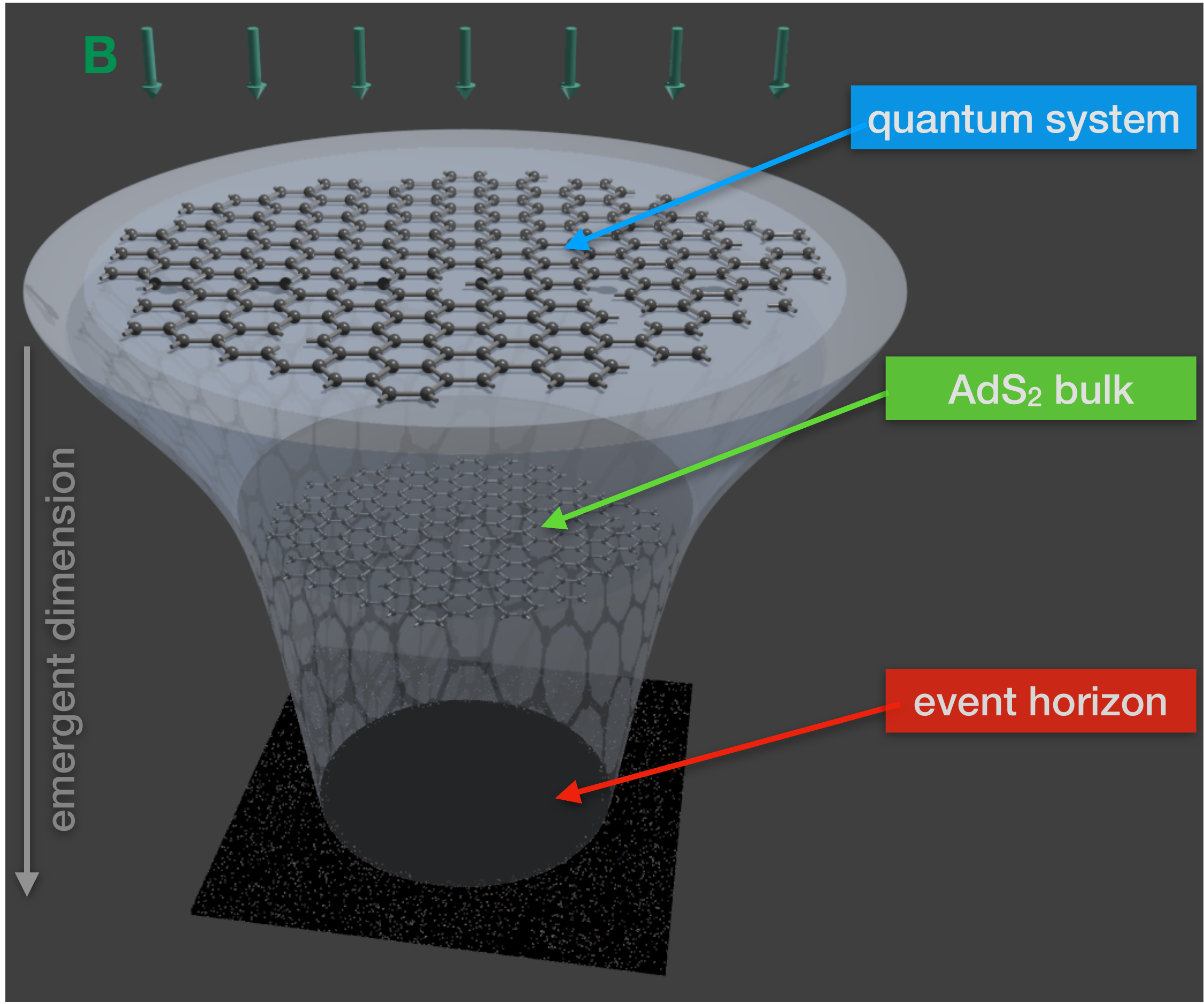}
\caption{{\bf Schematic representation of the holographic duality} 
 A (0+1) dimensional many-body system, represented here as a graphene
 flake in applied magnetic field, is holographically equivalent to a
  black hole in (1+1) anti-de Sitter space.
}\label{Fig1}
\end{figure}

Twenty years and many thousands of detailed calculations later, we now
have a very precise understanding of how gauge theories reproduce properties of quantum black holes immersed in AdS spaces. Note that astrophysical black holes are  well-described by classical physics, whereas (small, cold) quantum black holes perhaps play a role in the early evolution of the universe, but that avenue of investigation has many uncertainties. Through the gauge/gravity duality we now face the exciting prospect of accessing quantum black holes experimentally, in a new and extremely surprising context.

This precise correspondence and the possibility for experimental
realization raise an urgent question, namely: when does a quantum
many-body systems describe black holes in a theory of gravity? Which
aspects of the physics require a gravitational description? Or, put
more prosaically: given some material, what would one measure to
discover if it is  usefully described as a black hole in a (usually higher dimensional) gravitational theory?

\subsection{Many-Body Quantum Chaos}

The defining property of a black hole is its event horizon. To
identify quantum black holes in the lab, it is useful to identify
universal quantities, universal in that they only depend on the
structure of spacetime near the horizon. One such quantity was
identified by Shenker and Stanford \cite{Shenker2014} and later
employed by Kitaev, in an attempt to define a quantum mechanical analog of the butterfly effect and the associated Lyapunov exponent, which characterize that effect in classical chaos.

The classical butterfly effect quantifies the sensitivity to initial
conditions in chaotic systems. Nearby trajectories diverge
exponentially quickly from each other, with a rate that is called the
Lyapunov exponent. In quantum systems trajectories do not exist, but
one can still discuss the influence of an initial perturbation on
subsequent measurements. The suggestion was to quantify that
dependence by looking at the size of the commutator $[V(0),W(t)]$,
where the operator $V(0)$ provides the initial perturbation and $W(t)$
measures its influence at subsequent time $t$. Instead of calculating
the expectation value of $[V(0),W(t)]$, which defines the linear
response of the system, it turns out it is useful to focus on the
second moment: $C(t)={\rm tr}(e^{-\beta H} [V(0),W(t)]^2)$, which is
sometimes referred to as the {\em out-of-time-order correlator} or OTOC.  If we choose $V,W$ to be initially commuting Hermitian operators, the quantity $C(t)$ increases exponentially, as the information contained in the perturbation spreads rapidly throughout the system, a phenomenon referred to as ``scrambling". The exponent characterizing this exponential growth was dubbed the quantum Lyapunov exponent  $\lambda_L$. Various protocols for measuring the quantum Lyapunov exponent have been suggested.

Holographic quantum matter is in many ways as strongly interacting as
possible, consistent with quantum mechanical unitarity and causality
(or more precisely, limits on information propagation
\cite{Lieb:1972wy}). As such it is a useful arena to discover limits
on what is possible in highly quantum mechanical many-body
systems. There are many recent attempts to precisely quantify such
limits, which are typically saturated by holographic matter, starting
with the observation that the shear viscosity to entropy ratio for a
strongly interacting quantum liquid cannot become arbitrarily small
\cite{Kovtun:2004de}, an observation which played a role in
theoretical understanding of relativistic heavy ion
collisions. Perhaps the most precise and well-established such bound
to date is the bound on chaos: it is rigorously proven that the
quantum Lyapunov exponent obeys $\lambda_L \leq \frac{2 \pi}{\beta} $
for any quantum system \cite{Maldacena:2015waa}, where $\beta=1/k_BT$
is the inverse temperature.  Black holes are as strongly coupled as
possible, their Lyapunov exponents are as large as is allowed, and they scramble information as fast as is consistent with unitarity and causality.

Indeed, an elegant calculation in the gravitational context reveals
that the Lyapunov exponent depends only on the properties of black
holes near the event horizon, and is therefore universal in the sense
described above, that it takes the same values for all black holes in
the context of the AdS/CFT duality. Furthermore, that value is
precisely the maximally allowed value, $\lambda_L = \frac{2
  \pi}{\beta} $ for quantum black holes, realizing the intuition that
holographic matter is as strongly coupled as possible. This provides
an answer to the question posed in Box 1: looking at the manner in which information is scrambled in a quantum system can provide a connection to its gravitational description as a black hole. In particular, saturating the chaos bound is a strong indication of a connection to quantum black holes and gravitational description.

As it turns out, one can explicitly construct a simple model which has that property, that of maximal scrambling. This is  the so-called Sachdev-Ye-Kitaev (SYK) model, which we review presently.

\section{The SYK model and its large-$N$ solution}

\begin{figure}[t]
\includegraphics[width = 8.6cm]{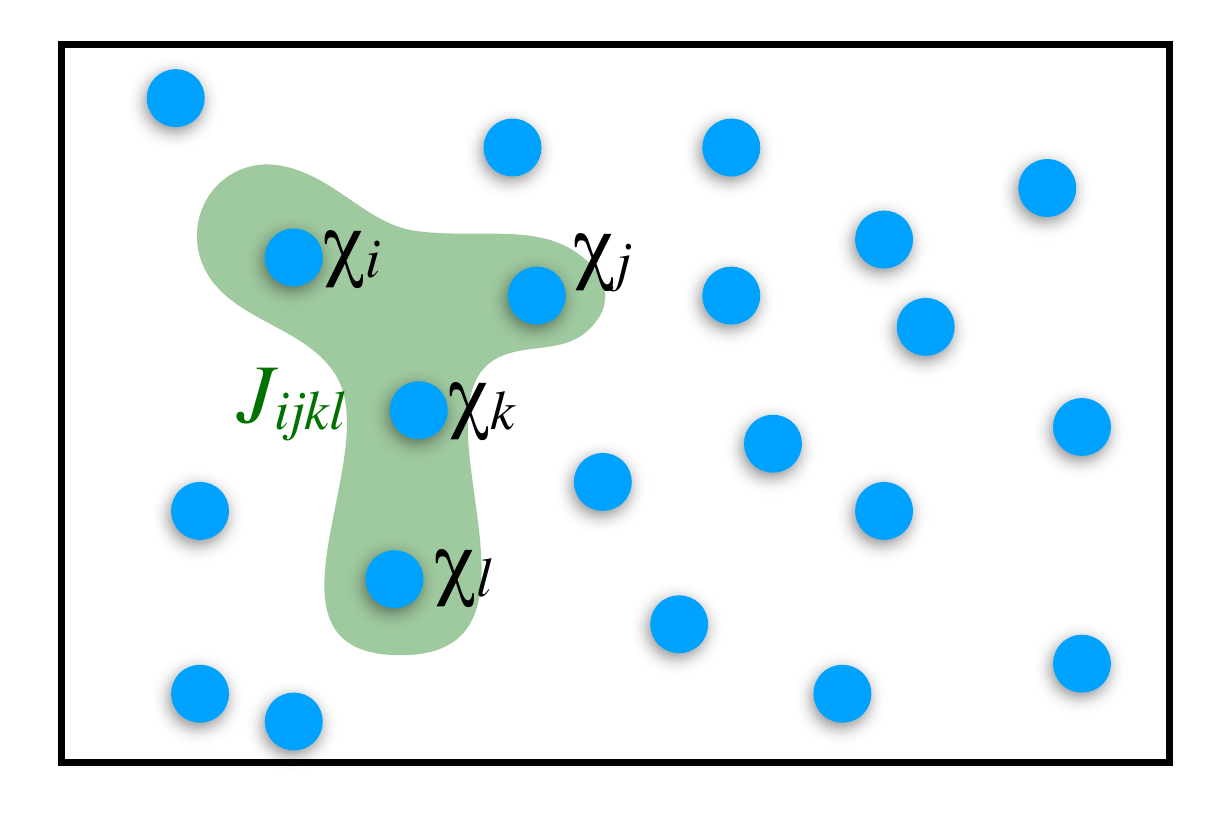}
\caption{{\bf The SYK model.}
   Blue dots represent Majorana fermions $\chi_j$ while the green
  region illustrates one possible 4-fermion interaction term in the Hamiltonian
  (\ref{hsyk}).
}\label{fig2a}
\end{figure}

The SYK model describes a system of $N$ interacting Majorana fermions. These are
particles identical to their antiparticles originally predicted in the
seminal work of Ettore Majorana \cite{Majorana1937,Wilczek2009,Alicea2012,Beenakker2012,Leijnse2012,Stanescu2013,Elliott2015} and recently observed
in a variety of solid-state platforms
\cite{Mourik2012,Das2012,Deng2012,Rokhinson2012,Finck2013,Hart2014,Nadj-Perge2014,Jia2016a,Jia2016b},
(also see Box 2). It is defined by the Hamiltonian 
\begin{equation}\label{hsyk}
{\cal H}_{\rm SYK}=\sum_{i<j<k<l} J_{ijkl}\chi_i\chi_j \chi_k\chi_l.
\end{equation}
schematically depicted in Fig.\ \ref{fig2a}.
Here $\chi_j$ with $j=1\cdots N$ represent the Majorana fermion operators that obey
the canonical anticommutation relations
\begin{equation}\label{hsyk2}
\{\chi_i,\chi_j\}=\delta_{ij}, \ \ \  \chi_j^\dagger=\chi_j.
\end{equation}
$J_{ijkl}$ are real-valued random independent variables drawn from
a Gaussian distribution with the width
\begin{equation}\label{J}
J^2={N^3\over 3!}\overline{J_{ijkl}^2}.
\end{equation}
The $N^3$ factor is necessary for the total energy of ${\cal H}_{\rm
  SYK}$ to scale extensively in the number of particles $N$, and for the interaction to stay finite in the large $N$ limit.  

Two key properties make  ${\cal H}_{\rm SYK}$ interesting and
non-generic: (i) the interactions between Majorana fermions are
all-to-all and completely random, and (ii) there is no term bilinear
in the fermion operators present in the Hamiltonian. The first property implies that there is no
concept of distance between the fermions and the Hamiltonian is therefore
zero-dimensional. By power counting it is easy to see that any
bilinear term in fermion operators would represent a relevant
perturbation to ${\cal H}_{\rm SYK}$. The second property thus ensures
that the model defined by Eq.\ (\ref{hsyk}) remains in the strong
coupling regime all the way to the lowest energies.

A closely related  model formulated  with complex fermions
\cite{French70,Bohigas71a,Bohigas71b}, also known as the complex
Sachdev-Ye-Kitaev (cSYK) model or  Sachdev-Ye (SY)
model  \cite{SY1996},  is defined  by the second-quantized Hamiltonian
\begin{equation}\label{hsy}
\cH_{\rm cSYK}=\sum_{ij;kl} J_{ij;kl}c^\dagger_ic^\dagger_jc_kc_l -\mu\sum_jc^\dagger_jc_j,
\end{equation}
where $c^\dagger_j$ creates a spinless complex fermion, $ J_{ij;kl}$ are
zero-mean complex random variables satisfying $ J_{ij;kl}=J^*_{kl;ij}$ 
and $J_{ij;kl}=-J_{ji;kl}=-J_{ij;lk}$ and $\mu$ denotes the chemical
potential. At half filling $(\mu=0)$ the two models exhibit very similar
behavior but  the cSYK model shows more complexity when $\mu\neq 0$. In this
review we will focus, for the most part, on the SYK model Eq.\
(\ref{hsyk}) as it is more straightforward to analyze. We comment on
the cSYK model as appropriate -- we will see that it might be easier to
realize it in a laboratory because it does not require the elusive
Majorana particles as basic building blocks. 

Perhaps the most remarkable and useful property of the SYK model is that, despite
being strongly interacting, it is exactly solvable
in the limit of large $N$. Specifically, it is possible to write down
a simple pair of equations for the averaged fermion propagator 
\begin{equation}\label{syk2}
G(\tau,\tau')={1\over N}\sum_j\langle {\cal T}_\tau\chi_j(\tau)\chi_j(\tau')\rangle
\end{equation}
and the corresponding self energy $\Sigma(\tau,\tau')$, that become
asymptotically exact in the limit $N\to \infty$ and have a simple
solution in the low-frequency limit. This solution is reviewed in Box
3 and leads to some remarkable conclusions.

The equations governing the large-$N$solution (\ref{syk4})  display an intriguing
time-reparametrization (also referred to as conformal) invariance at low energies which hints at a
connection of the model to black holes and string theory.  
As a practical matter the invariance 
allows one to extract the low-frequency behavior of the propagator,
\begin{equation}\label{prop6}
G_c(\omega_n)=i\pi^{1/4}{\sgn(\omega_n)\over\sqrt{J|\omega_n|}}
\end{equation}
and the corresponding spectral function 
\begin{equation}\label{prop7}
A_c(\omega)={1\over \sqrt{2}\pi^{3/4}}{1\over \sqrt{J|\omega|}},
\end{equation}
where subscript $c$ denotes the conformal regime
$|\omega|\ll J$. At high frequencies the behavior must cross over to
$1/\omega$ in both cases.  The absence of a pole in $G_c(\omega_n)$
indicates the expected non-Fermi liquid behavior of the SYK model. In
addition we will see that the
characteristic inverse square root singularity of the spectral
function could be measurable by various spectroscopies in some of the
proposed experimental realizations of the model. 

The conformal structure implied by Eqs.\ (\ref{syk4}), has also
been instrumental in extracting the quantum-chaotic properties of the
SYK model. It allows for the calculation of the out-of-time-order
correlator $C(t)$ which shows the characteristic exponential growth 
\cite{Kitaev2015,Maldacena2016}
with the maximal permissible Lyapunov exponent $\lambda_L = \frac{2
  \pi}{\beta}$, thus confirming the intuition that the SYK model is indeed
holographically connected to a black hole.

Finally we mention various important extensions of the SYK model
developed in the recent literature. These include models
showing unusual  spectral properties  \cite{Xu2016,Polchinski2016,Verbaar2016}, supersymmetry  \cite{Fu2016}, quantum phase transitions of an
unusual type  \cite{Altman2016, Bi2017,Lantagne2018}, quantum chaos propagation
 \cite{Gu2016,Berkooz2016,Hosur2016}, and patterns of entanglement \cite{Liu2017,Huang2017}.

\subsection{Box 2: Majorana zero modes and the Kitaev chain}

Majorana fermions have been originally introduced in the context of
particle physics as special solutions of the celebrated Dirac equation \cite{Dirac1928}
describing relativistic quantum mechanics of spin-${1\over 2}$
particles.  Unlike ordinary fermions (e.g. electrons or protons) Majorana
particles lack the distinction between particle and antiparticle
\cite{Majorana1937}. In the elementary particle physics the leading
candidate for a Majorana fermion is the neutrino but the experimental
evidence remains inconclusive at present \cite{Elliott2015}. 

\begin{figure}[t]
\includegraphics[width = 8.6cm]{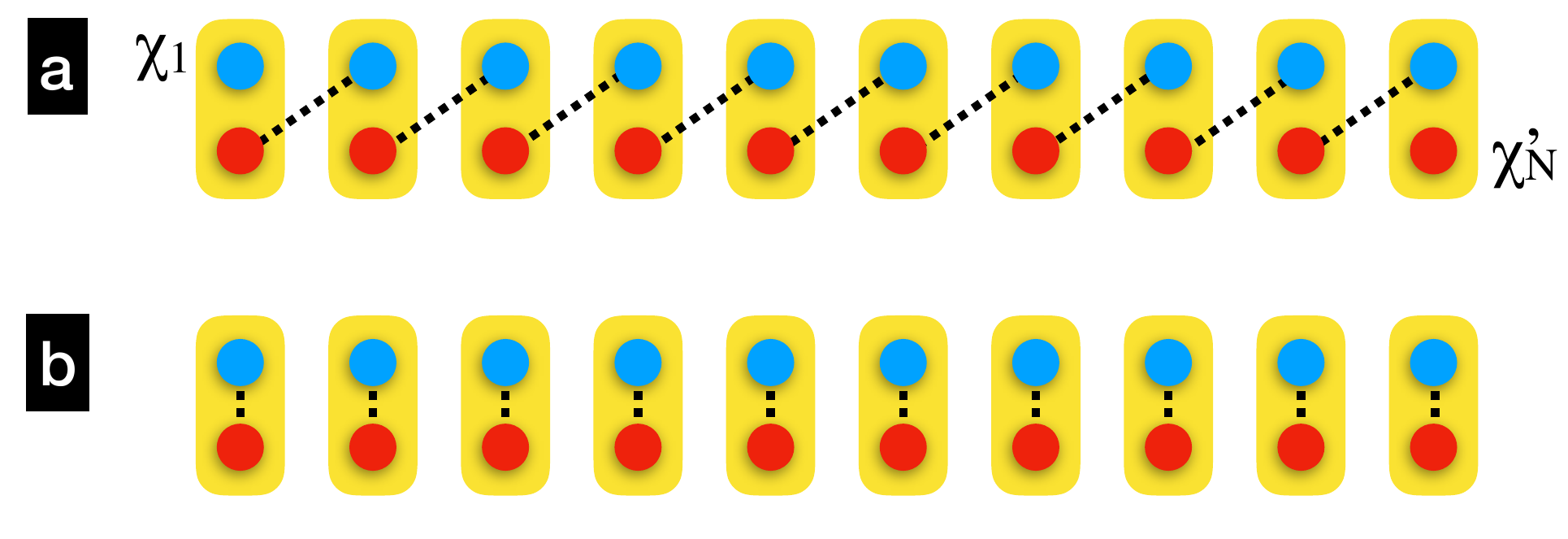}
\caption{{\bf The Kitaev chain.} 
Blue and red dots represent Majorana modes $\chi_j$ and $\chi_j'$
associated with site $j$ of the chain.
  a) In the topological phase a bond forms between $\chi_j'$ and
  $\chi_{j+1}$ for each $j$, leaving $\chi_1$ and $\chi_N'$ unpaired. 
b) In the trivial phase  $\chi_j$ and $\chi_j'$ pair up on the each
site leaving behind no unpaired end modes.  
}\label{fig3}
\end{figure}
In condensed matter physics Majorana fermions can appear as emergent
particles in various many-body systems. The most prominent of these
are topological superconductors in which Majorana particles often
appear as topologically protected zero modes
\cite{Alicea2012,Beenakker2012,Leijnse2012,Stanescu2013,Elliott2015}. A
canonical example of such a topological superconductor is the Kitaev chain
\cite{Kitaev2001} which we now briefly review to illustrate the
emergence of Majorana zero modes in a simple setting. 

Consider spinless fermions in a 1D lattice. The simplest model of a
superconductor in such a setting is described by the Hamiltonian 
\begin{equation}\label{hkit1}
{\cal H}=\sum_j\biggl[-t (c^\dagger_j c_{j+1}+{\rm h.c.}) - \mu  c^\dagger_j c_{j}
+ (\Delta c^\dagger_j c^\dagger_{j+1}+{\rm h.c.})\biggr],
\end{equation}
where $c^\dagger_j$ creates a fermion on site $j$ and satisfies the
usual fermionic anticommutation relation 
\begin{equation}\label{hkit2}
\{c^\dagger_i,c_j\}=\delta_{ij}.
\end{equation}
$t$ and $\Delta$ represents the nearest-neighbour hopping and  pairing amplitudes,
respectively and $\mu$ denotes the chemical potential. We assume that
$\Delta$ is real and consider a chain with $N$ sites and open boundary
conditions.  

Following Kitaev \cite{Kitaev2001} we now express each Dirac fermion $c_j$ in
terms of two Majorana fermions $\chi_j$ and $\chi_j'$,
\begin{equation}\label{hkit3}
c_j={1\over\sqrt{2}}(\chi_j+i\chi_j'), \ \ \ c_j^\dagger={1\over\sqrt{2}}(\chi_j-i\chi_j').
\end{equation}
It is easy to check that if the Majorana operators obey the
anticommutation algebra given by Eq.\ (\ref{syk2}) then Eq.\
(\ref{hkit2}) is satisfied and the transformation is thus
canonical. In the Majorana representation the Hamiltonian (\ref{hkit1})
becomes
\begin{equation}\label{hkit4}
{\cal H}=i\sum_j\biggl[(\Delta+t)\chi_{j}'\chi_{j+1}
+(\Delta-t)\chi_{j}\chi_{j+1}' -\mu \chi_{j}\chi_{j}' \biggr].
\end{equation}
To appreciate the physical content of this Hamiltonian it is useful to
consider a special limit in which $\mu=0$ and $\Delta=t$. We then have simply
\begin{equation}\label{hkit5}
{\cal H}=2it\sum_{j=1}^{N-1}\chi_{j}'\chi_{j+1},
\end{equation}
where we restored the summation bounds. 

Equation (\ref{hkit5}) is remarkable for what it is lacking:
notice that two operators, $\chi_1$ and $\chi_N'$, do not participate
in the Hamiltonian. They therefore constitute exact ``zero modes'' of
the system. If we construct a new Dirac fermion operator
$\tilde{c}=(\chi_1+i\chi_N')/\sqrt{2}$ then it is clear that it costs
zero energy to add such a $\tilde{c}$-particle to the system. Remarkably,
the zero mode is fundamentally delocalized between the two ends of the
chain as illustrated in figure \ref{fig3}. Kitaev \cite{Kitaev2001}  noted
the possibility of storing quantum information in this delocalized
state which would make it resilient to many sources of
decoherence. Such a possibility also underlies much of the current
interest in Majorana zero modes.

With a little extra work one can demonstrate that he special limit
discussed above represents merely a point in the stable topological phase of
the Hamiltonian (\ref{hkit1}). The system exhibits gapped bulk with two
Majorana zero modes localized near the ends of the chain whenever
$\Delta\neq 0$ and $|\mu|<2|t|$. When $|\mu|>2|t|$ the chain is
topologically trivial and zero modes are absent. Various instances of Majorana zero
modes observed experimentally in 1D systems
\cite{Mourik2012,Das2012,Deng2012,Rokhinson2012,Finck2013,Hart2014,Nadj-Perge2014}
can all be understood as realizations of the Kitaev chain.

\subsection{Box 3: The large-$N$ solution}
The fermion propagator of the SYK model  (\ref{syk2}) is related to the self energy by 
the standard Dyson equation, 
\begin{equation}\label{syk3}
G(\omega_n)=[-i\omega_n-\Sigma(\omega_n)]^{-1}, 
\end{equation}
where we have assumed time-translation invariance and defined 
$G(\omega_n)=\int_0^\beta d\tau e^{i\omega_n\tau}G(\tau)$, with Matsubara frequencies $\omega_n=\pi T(2n+1)$
and $n$ integer.
The solution exploits the fact that for large $N$ diagrammatic
expansion of the self energy $\Sigma(\tau,\tau')$  is dominated by a class of the so called
melonic diagrams which can be summed up exactly. All other diagrams,
including the vertex corrections that typically prevent such exact
resummations, are suppressed by factors of $1/N$ and can be neglected.
\begin{figure}[t]
\includegraphics[width = 8.6cm]{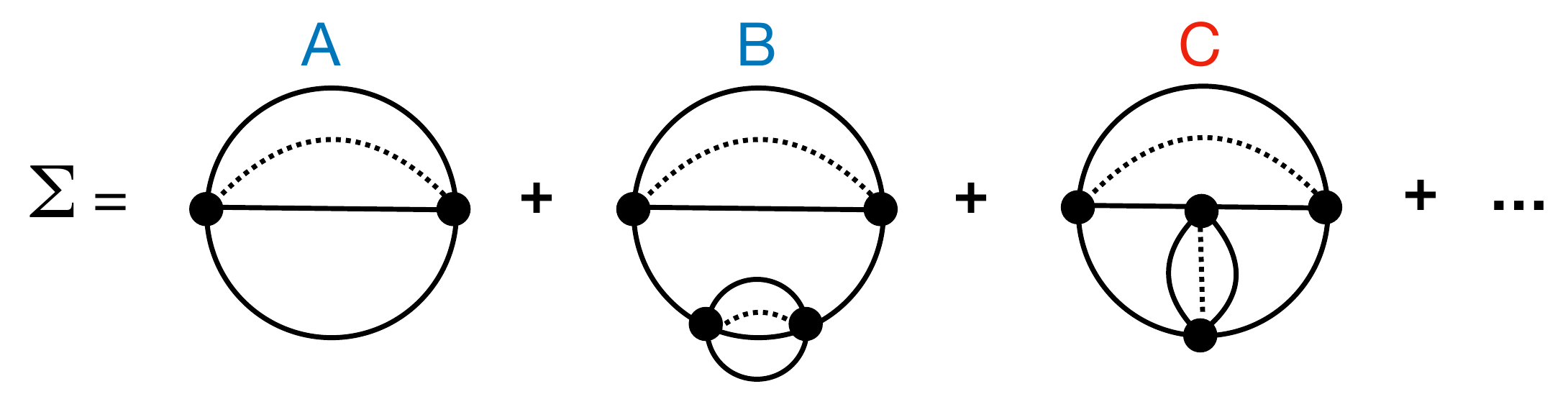}
\caption{{\bf The large-$N$ solution.}
Leading Feynman diagrams in the $1/N$ expansion of the self energy
$\Sigma(\tau)$. Solid lines represent the fermion propagator $G(\tau)$
while dashed lines indicate disorder averaging over the pairs of
vertices connected by the line.  
}\label{fig2b}
\end{figure}

This can be seen from Fig.\ \ref{fig2b} which illustrates various
diagrams that contribute
to $\Sigma$. The leading melonic diagram A has two vertices.
Upon averaging over disorder Eq.\ (\ref{J}) implies that  these contribute a factor
$\overline{J_{ijkl}^2}\sim J^2/N^3$. Summing over the three internal legs contributes a factor
$\sim N^3$ as each leg can represent one of the $N$ fermion flavours. The
diagram is therefore $O(1)$. It is easy to check that all diagrams
with melonic insertions, such as the  diagram B are
$O(1)$. On the other hand diagrams representing vertex corrections,
such as diagram C, have fewer internal legs and
therefore become $O(1/N)$ and thus negligible at large $N$.  

It is straightforward to sum up the $O(1)$ diagrams comprising the
self energy -- the full series is generated by simply replacing the
bare propagator  $G_0(\omega_n)=-1/i\omega_n$ by the fully dressed
propagator $G(\omega_n)$ in the leading melonic diagram A in Fig.\ \ref{fig2b}. One thus obtains
\begin{equation}\label{syk3b}
\Sigma(\tau)=J^2G^3(\tau),
\end{equation}
which together with Dyson equation (\ref{syk3}) determine the form of
the  propagator at large $N$.

The mean field equations for the bi-local fields $G, \Sigma$ become
exact in the large $N$ limit, where we can ignore fluctuations around
their saddle point values, which are the solution of the Schwinger-Dyson
equations. The property of being mean-field exact is shared by many
infinite-ranged models, for example the celebrated
Sherrington-Kirkpatrick model of spin glasses \cite{SK1975}.  Remarkably, the low energy physics of the SYK model is very different from most such models, as it avoids replica symmetry breaking and instead stays quantum mechanical to the lowest energies.

At sufficiently low frequencies $\omega_n\ll J$ the $i\omega_n$ term in Eq.\
(\ref{syk3}) can be neglected and the resulting system of two
equations can be seen to acquire time parametrization invariance
\begin{equation}\label{syk4}
\begin{split}
G(\tau,\tau')&\to [f'(\tau) f'(\tau')]^{1/4}G(f(\tau),f(\tau')), \\
\Sigma(\tau,\tau')&\to  [f'(\tau)
                    f'(\tau')]^{3/4}\Sigma(f(\tau),f(\tau')), 
\end{split}
\end{equation}
for an arbitrary smooth function $f(\tau)$. This emergent conformal
symmetry is a key property of the SYK model which hints at its deep
connection to black hole physics and holography.  In fact, it is described by the same low energy action (the so-called Schwarzian action) as black holes in asymptotically AdS$_2$ \footnote{The connection between the cSYK models and gravity in AdS$_2$ was first pointed out in \cite{Sachdev:2010um}.}. These low-dimensional black holes arise as the near-horizon region of a large class of charged black holes in 3+1 dimensions (see e.g \cite{Nayak:2018qej}), thus the Schwarzian action also describes the low energy dynamics of those more realistic black holes.

The relation to black hole physics is explained in greater detail in \cite{Maldacena:2016upp} and many subsequent works. We refer the interested reader to that vast literature for the many fascinating details of that correspondence.

\section{Proposed physical realizations}

Given the plethora of interesting phenomena that are thought to be
encoded in the SYK model and the rich connections it furnishes between
disparate fields of physics, it would be of obvious interest to have
its experimental realization. It has been proposed that high-$T_c$
cuprate superconductors and some other strongly correlated ``strange
metals'' may realize various higher-dimensional generalizations of the SYK
model \cite{Gu2017, Davidson2017,Balents2017,Zhang2017,Patel2018,Chowdhury2018,Lawler2018}. This connection is made on phenomenological grounds and
although very appealing, it remains speculative because of the 
difficulties encountered in the attempts to relate microscopic models
governing crystalline solids to highly random SYK model (see however
Ref.\ \cite{Wu2018} for an SYK-like model without randomness). 

Another class of interesting approaches are quantum simulations. An
algorithm has been proposed for digital quantum simulation of both SYK
and cSYK models using a quantum computer \cite{Garcia2017}. First
experimental steps along this line have already been reported \cite{Laflamme2017}. 

In this
review we focus on proposals that aim at engineering the cSYK or SYK model
bottom up, that is, by designing systems with requisite degrees of
freedom and interactions using building blocks that are reasonably well
understood and thus afford a high degree of control.

\subsection{Three key challenges}

When thinking about the experimental realization of the SYK model several
challenges become immediately apparent. The original SYK model employs
Majorana fermions so the first challenge is to design a physical
system capable of hosting a large number $N$ of these elusive
particles. In the language of condensed matter physics one needs
Majorana zero modes
\cite{Alicea2012,Beenakker2012,Leijnse2012,Stanescu2013,Elliott2015}
which have been observed to exist in several platforms including
semiconductor quantum wires
\cite{Mourik2012,Das2012,Deng2012,Rokhinson2012,Finck2013}, atomic
chains \cite{Nadj-Perge2014}, and vortices in topological insulator -
superconductor interfaces \cite{Jia2016a,Jia2016b}. In these works
signatures consistent with individual localized Majorana modes have
been reported but manipulating and assembling them into required
multi-mode structures remains a considerable challenge. 

Once we have
an assembly of $N$ Majorana particles the second challenge arises. The
leading perturbation in such a situation, typically, will come from the
hybridization term 
\begin{equation}\label{heff2}
{\cal H}_{2}={i}\sum_{i<j} K_{ij}\chi_i\chi_j
\end{equation}
which is quadratic in Majorana operators, not quartic as required for the
SYK model. In Eq.\ (\ref{heff2}) real-valued constants $K_{ij}$ arise
from the overlap between Majorana fermion wavefunctions and describe
tunnelling events much like overlaps between atomic orbitals in
molecules or solids which describe electron tunnelling. In the presence of
interactions (such as those arising from the Coulomb repulsion between
the underlying electron degrees of freedom) the full Hamiltonian
of the system of $N$ Majorana particles will be $\cH=\cH_2+\cH_{\rm
  SYK}$. For such a Hamiltonian it is known \cite{Altman2016} that $\cH_2$ is a relevant
perturbation to $\cH_{\rm SYK}$ meaning that for arbitrary non-zero
$K_{ij}$ the SYK fixed point is destroyed and the ground state of the
system becomes a disordered Fermi liquid. Fortunately, vestiges of the
SYK physics still survive in the excited states above the crossover
energy scale $E_c\simeq K^2/J$ where we have defined the average
hybridization strength as    
\begin{equation}\label{K}
K^2={N}\overline{K_{ij}^2}.
\end{equation}
The excited states can be probed by performing experiments at non-zero
temperature $T$ or frequency $\omega$, such that $K^2/J <T,\omega <
J$. In order to have a reasonably wide range of temperatures or
frequencies in which the SYK physics remains observable one therefore
needs to ensure that $K\ll J$. This is the second key challenge
facing any experimental realization of the SYK model as a generic
system of $N$ Majorana modes would instead have $K\gg J$.

The final challenge has to do with ensuring sufficient randomness in
coupling constants $J_{ijkl}$. In order to define the SYK Hamiltonian
these constants must be sufficiently close to random independent
variables with a Gaussian distribution. As we shall see such
randomness can arise from the microscopic disorder that is inherently
present in solids or else  it must be deliberately
engineered in the system through fabrication. 

Finally we remark that the first of the challenges discussed above is avoided in proposals
that aim at realization of the cSYK model Eq.\ (\ref{hsy}) which employs
ordinary complex fermions as basic building
blocks and not Majorana fermions. Indeed these proposals appear to be
simpler to implement and are therefore likely to succeed first. As we mentioned
the cSYK model at half filling exhibits the same holographic duality to a black hole
as the SYK model and it shows additional interesting behavior away
from half filling \cite{Sachdev2015}. In solid state systems
complex fermions will be electrons and will thus carry
electric charge. For this reason such systems will be amenable to
standard transport and spectroscopic measurements. This confers another
advantage over Majorana fermions, which are electrically neutral.      

\subsection{cSYK model with ultracold gases}

Experiments on ultracold atomic gases in optical lattices have already
succeeded in realizing a number of theoretical models originally introduced in
the context of condensed matter physics. Notable examples include the
Bose-Hubbard model \cite{Greiner2002} and the Haldane model
\cite{Jotzu2014}. Owing to their high degree of controllability,
lack of disorder, and the unique set of physical observables
experiments on ultracold gases can often answer many important
questions about these models that remain inaccessible to their
condensed  matter analogs.

\begin{figure}[t]
\includegraphics[width = 7.6cm]{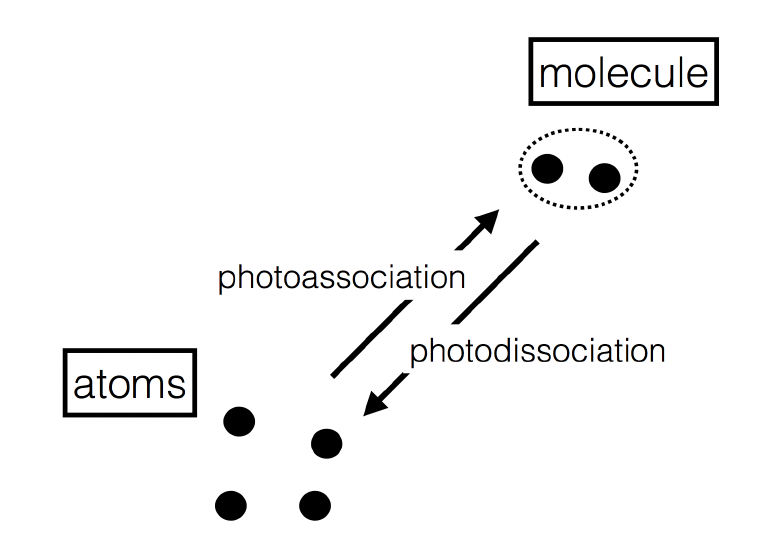}
\caption{{\bf cSYK model from cold atoms.}
Laser beams are used to couple atoms and molecules such that
transitions occur as described by Hamiltonian (\ref{dht1}.
(Figure copied from Ref.\ \cite{Danshita2017b}).
}\label{fig4}
\end{figure}

A proposal to realize the cSYK model with ultracold  gases in optical
lattices has been given by Danshita, Hanada and Tezuka 
\cite{Danshita2017,Danshita2017b}. The authors envision fermionic atoms, such as
$^6$Li, confined in deep potential wells of an optical lattice. Each
well is assumed to have $N$ strongly localized atomic states filled with $Q$
atoms. While the eigenstate wavefunctions spatially overlap in the same well,
their orthogonality forbids single particle tunnelings, so that
perturbations similar to Eq. (4.1), bilinear in fermion operators, 
can be ignored. The atoms in separate wells are also assumed to be completely
decoupled from each other so we can focus on a single well. In addition, two atoms can
come together and form bosonic molecular states through the process of
photo-association (PA) stimulated by PA lasers of appropriate
frequencies, as illustrated in Fig.\ \ref{fig4}.  The Hamiltonian governing  such a system is
\begin{eqnarray}
\cH&=&\sum_{s=1}^{n_{\rm ms}}\biggl[\nu_sm_s^\dagger m_s+
\sum_{s'=1}^{n_{\rm ms}}{U_{ss'}\over 2} m_s^\dagger m_{s'}^\dagger
    m_{s'}m_s \nonumber \\
&+&\sum_{i,j} g_{s,ij}\left(m_s^\dagger c_ic_j-m_s c_i^\dagger c_j^\dagger\right)\biggr],
\label{dht1}
\end{eqnarray}
where $m_s^\dagger$ and $c_i^\dagger$ represent the creation operators
for molecules and atoms, respectively, $\nu_s$ are molecular state
energies, $U_{ss'}$ are molecular interactions and $g_{s,ij}$ denote the
amplitudes of PA processes.  The latter are assumed fully controlable
as they depend on frequencies and intensities of the PA lasers.

The molecular states can be integrated out from the model defined by
Eq.\ (\ref{dht1}) and, using perturbation theory to second order in amplitudes $g_{s,ij}$, one
obtains the effective Hamiltonian for the fermions of the form
\begin{equation}\label{dht2}
\cH_{\rm eff}=\sum_{i,j,k,l}\left(\sum_{s=1}^{n_{\rm ms}} {g_{s,ij}g_{s,kl}\over\nu_s}\right) c_i^\dagger c_j^\dagger c_kc_l.
\end{equation}
This clearly has the structure of the cSYK model Eq.\ (\ref{hsy})
provided that we identify the sum in the brackets with $J_{ij;kl}$.
The coupling constants in this realization of the cSYK model are real
(as opposed to complex-valued) but the authors \cite{Danshita2017}
argue convincingly that for large $N$ this difference does not
matter. Also, unlike the canonical cSYK model,  $J_{ij;kl}$ defined by
Eq.\ (\ref{dht2}) have internal structure and are not in general Gaussian distributed random
variables.  If however  $g_{s,ij}$ and $\nu_s$ are taken as random variables
then for a large number of molecular states $n_{\rm ms}\gg 1$ the
distribution of $J_{ij;kl}$ should approach Gaussian by virtue of the
central limit theorem ($J$s are then given by a sum of a large number
of identically distributed random variables). In this limit, at least,
the system defined by Hamiltonian (\ref{dht1}) should behave as a
canonical cSYK model. 

Unfortunately,  $n_{\rm ms}\gg 1$ is also the limit which might be
difficult to realize in a laboratory. The chief bottleneck is the
number of lasers required to drive the PA transitions between the
atomic and molecular states. This equals $n_{\rm ms} N(N-1)/2$ which
becomes a very large number in the interesting limit when both $n_{\rm
  ms}$ and $N$ are large. In addition, frequencies of these PA lasers must be rather
accurately tuned to achieve reasonable rates of transitions. It is
known from numeical studies that $N\gtrsim 10$ is required to start observing spectroscopic
signatures characteristic of the cSYK model. It is less clear how large
$n_{\rm ms}$ must be for a given $N$ to yield
Gaussian distributed $J$s to a good approximation. The cSYK model variant defined by Eq.\ (\ref{dht2}) could however
exhibit interesting behavior even for small $n_{\rm ms}$  and a
theoretical study of this limit would be of interest.

The authors \cite{Danshita2017}, among other things, devised a protocol to measure the
out-of-time-order correlator $C(t)$ in this sytem, following the
general procedure outlined in Ref.\ \cite{Swingle2016}. The protocol involves
coupling the atomic states to a control qubit, annihilating individual
atoms at different times according to the qubit state, and finally
evolving the whole system forward and backward in time according to
the Hamiltonian (\ref{dht2}). Backward time evolution requires
flipping the sign of the Hamiltonian which is achieved by
flipping the sign of all $\nu_s$. This in turn can be implemented by
changing the detuning of the PA lasers. Measuring OTOC is clearly a
challenging proposition but first steps have recently been taken using
quantum simulators \cite{Garttner2017,Li2017}.

\subsection{cSYK model in a graphene flake}

This proposed solid-state realization \cite{achen2018} takes its inspiration from
the rich physics of quantum Hall liquids \cite{Tsui1982,Laughlin1983}. It is well known that
application of a strong
magnetic field to the 2D electron gas (electrons confined to a thin layer) 
results in quenching of their quantum kinetic energy and strong
enhancement of the interaction effects.  This is equivalent to the
suppression of two-fermion terms  $\cH_2$ relative to the four-fermion
interaction terms contained in $\cH_{\rm cSYK}$.  In a large clean sample one
then obtains the family of fractional quantum Hall liquids
characterized by a range of spectacular physical properties including
topological order as well as the charge and statistics
fractionalization \cite{Arovas1984}.
\begin{figure}[t]
\includegraphics[width = 7.6cm]{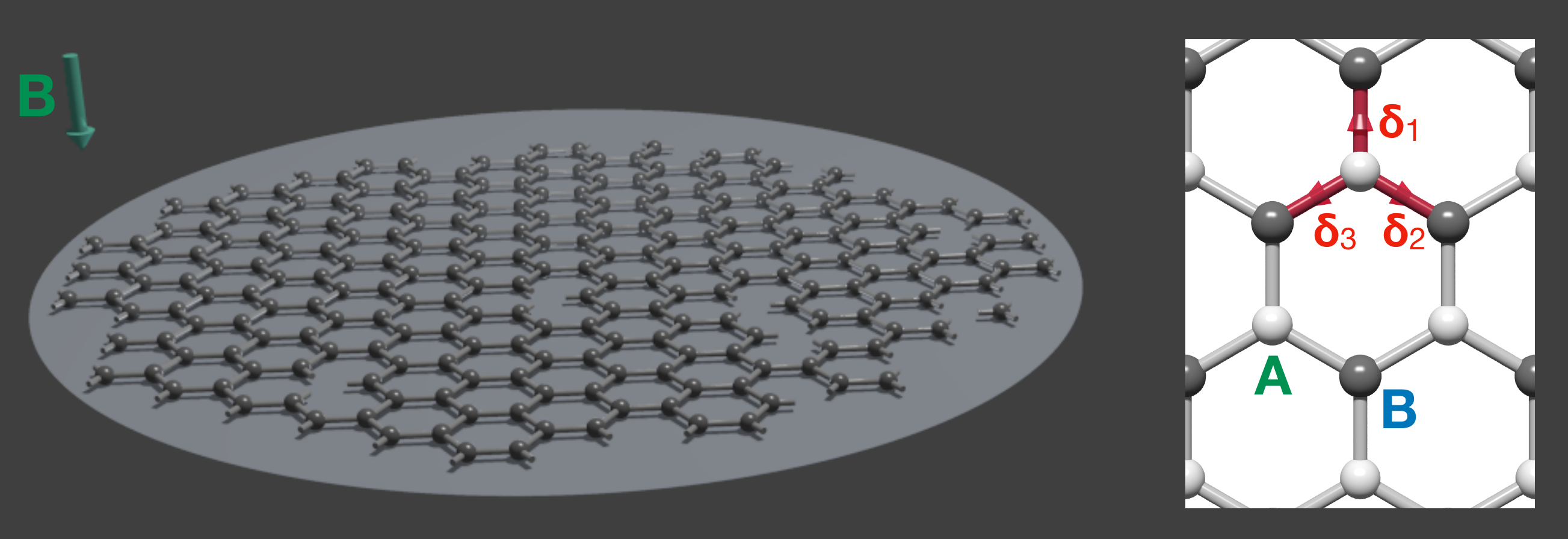}
\caption{{\bf Schematic representation of the graphene device.}
  Irregular shaped graphene flake in applied magnetic field 
  $B$ hosts degenerate Landau levels which can serve as a platform for
  the cSYK model when interactions are present. Inset: lattice structure of
  graphene with A and B sublattices marked and nearest neighbour
  vectors denoted by $\bdel_j$. (Figure copied from Ref.\ \cite{achen2018}).
}\label{fig5}
\end{figure}

The root cause for the above behavior is Landau quantization: at the
non-interacting level magnetic field reorganizes the electron bands into a
discrete set of massively degenerate dispersionless Landau levels (LLs). The
effective single-particle Hamiltonian for an electron in a given
Landau level is then simply $\cH_2=0$. The proposal of Ref.\
\cite{achen2018} uses the Landau level as the starting point to
construct the cSYK model. The key remaining challenge, then, is to produce
coupling constants $J_{ij;kl}$ that would be all-to-all
and sufficiently random. To achieve this one must include disorder
effects. Under normal circumstances, however, microscopic disorder tends to
broaden LLs, effectively reintroducing nonzero $\cH_2$. For this
reason ordinary 2D electron gas with disorder does not work as a
platform to build the cSYK model. 

The idea is to use instead a flake of graphene with a highly irregular
boundary serving  as a source of disorder, see Fig.\ \ref{fig5}. Electrons in graphene are well-known
to posses ``relativistic'' energy dispersion described at low energies
by a massless Dirac Hamiltonian \cite{Neto_RMP}. Landau levels of such
a Hamiltonian occur at energies 
\begin{equation}\label{flake1}
E_n\simeq\pm\hbar v\sqrt{2n(eB/\hbar c)}
\end{equation}
with the characteristic Fermi velocity $v$  and $n=0,1,\cdots$. The
$n=0$ Landau level, often referred to as LL$_0$, is special in terms
of how it responds to disorder. According to the famous
Aharonov-Casher construction \cite{Casher1979} the exact degeneracy of
quantum states in LL$_0$ is protected against any disorder that respects the
chiral symmetry of graphene. This is most simply explained by
inspecting the tight-binding model describing electrons in the
graphene  honeycomb lattice, 
\begin{equation}\label{flake2}
H_0=-t\sum_{\br,\bdel} (a^\dagger_\br b_{\br+\bdel}+{\rm h.c.}).
\end{equation}
Here $a^\dagger_\br\ (b^\dagger_{\br+\bdel})$ denotes the creation
operator of the electron on the sublattice  A (B)  of the
honeycomb lattice. $\br$ extends
over the sites in sublattice A while $\bdel$ denotes the 3 nearest
neighbour vectors (inset Fig.\ \ref{fig5}),   $t=2.7$ eV is the
nearest-neighbour tunneling amplitude.  
The chiral symmetry $\chi$ is generated by setting
$(a_\br,b_\br){\rightarrow} (-a_\br,b_\br)$ for all $\br$ which has the effect of flipping the
sign of the Hamiltonian $H_0\to -H_0$. 

Making the shape of the flake irregular introduces randomness that
respects $\chi$. The wavefunctions $\Phi_j(\br)$ of the states belonging to LL$_0$ acquire
random spatial structure but, importantly,  Aharonov-Casher
construction \cite{Casher1979}  guarantees that their energies remain perfectly
degenerate. Under these circumstances one expects the Coulomb
interaction $V(r)$ to produce random and all-to-all coupling constants
$J_{ij;kl}$ between the electrons in LL$_0$. This is because
$J_{ij;kl}$ are given as matrix elements of $V(r)$ between random-valued
wavefunctions $\Phi_j(\br)$ which, according to simulations
\cite{achen2018}, have support everywhere in the flake.

Characteristic signatures of the cSYK physics could be observed in this
system by a host of spectroscopic and transport probes. Scanning
tunneling spectroscopy measures the local density of states which is directly
related to the spectral function and should thus exhibit the
characteristic inverse square root divergence Eq.\ (\ref{prop7}) as a
function of bias voltage. Two-terminal conductance through the system
and local compressibility, measured via charge sensing techniques,
should likewise show signatures of the non-FL behavior.  It is
currently not known how to measure out-of-time-ordered quantities, such as $C(t)$ in
this context and this remains a challenge for all solid-state
realizations of cSYK and SYK models. 

\subsection{SYK model with semiconductor quantum wires}

Semiconductor quantum wires with strong spin-orbit coupling, such as
those made of InSb, have emerged as the prime platform for investigating
Majorana zero modes. When such wires are ``proximitized'' (that is,
made superconducting by placing them in contact with a superconducting
material, such as Al) and placed in magnetic field,  Majorana zero modes are predicted to appear at
their ends \cite{Oreg2010,Lutchyn2010}. There now exists ample
experimental evidence for this effect
 \cite{Mourik2012,Das2012,Deng2012,Rokhinson2012,Finck2013,Hart2014}
 which realizes the Kitaev chain paradigm discussed in Box 2.  
\begin{figure}[t]
\includegraphics[width = 7.6cm]{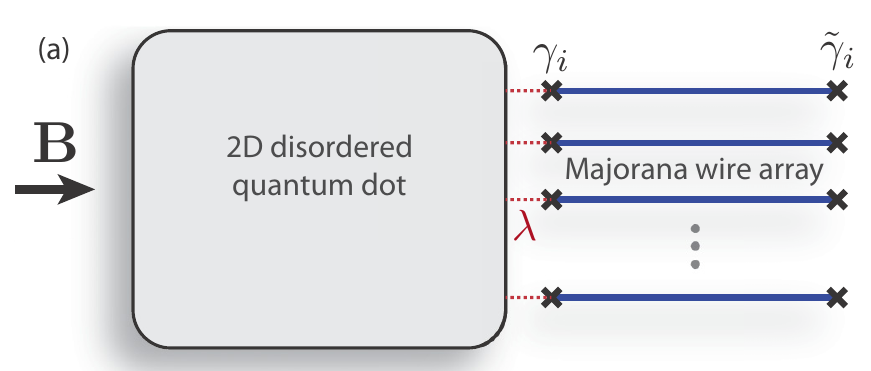}
\caption{{\bf SYK model with quantum wires.}
An array of quantum wires with Majorana end modes is coupled to a
disordered quantum dot. Majorana modes penetrate into the dot and
interact with one another thus realizing the SYK model. 
(Figure copied from Ref.\ \cite{Chew2017}).
}\label{fig6}
\end{figure}

Chew, Essin and Alicea \cite{Chew2017} proposed to engineer the SYK
Hamiltonian by coupling $N$ such wires to a disordered 2D  quantum dot
as illustrated in Fig.\ \ref{fig6}. In such a setup Majorana zero modes can be
shown to spread out from the tips of the individual wires into the quantum
dot so that their wavefunctions are essentially delocalized across
its entire area. Once delocalized Majorana wavefunctions overlap in real
space and even perfectly local interactions produce all-to-all matrix
elements $J_{ijkl}$ precisely as required by the SYK Hamiltonian. The fact that
the quantum dot is disordered means that the wavefunctions acquire
random spatial structure which in turns translates to randomness in $J$s. 

Two features of this proposed setup deserve note. First, Majorana zero
modes avoid hybridization (that is, formation of bilinear terms in the
Hamiltonian that would be detrimental to SYK) by virtue of approximate artificial time-reversal
symmetry ${\tilde\cT}$ present in the system.  This may be illustrated
most easily on the example of the Kitaev chain. Its Hamiltonian Eq.\
(\ref{hkit1}) respects an antiunitary symmetry ${\tilde\cT}$ which
sends $c_j\to c_j$ and $i\to -i$. In the Majorana representation  this
symmetry is expressed as 
\begin{equation}\label{dot1}
{\tilde\cT}: \ \ \chi_j\to \chi_j, \ \ \ \chi'_j\to -\chi'_j, \ \ \
i\to -i,
\end{equation}
and it is easy to check that Hamiltonian (\ref{hkit3}) is indeed
invariant. If we now assume that from each of $a=1\dots N$ wires only the left-most
Majorana mode $\chi_1^{a}$ is coupled to the dot, and that the entire
system continues to respect ${\tilde\cT}$, then bilinear terms of the
form $iK_{ab}\chi_1^{a}\chi_1^{b}$ are prohibited as they would be odd
under the symmetry. Of course ${\tilde\cT}$  is not a true microscopic
symmetry of the physical system so strictly speaking $K_{ab}$ is not
zero. Ref. \ \cite{Chew2017} however argued that it is an approximate
symmetry which implies that $K_{ab}$ are likely to be small compared to the
interaction energy scale. 

The second feature has to do with the hybridization of the Majorana modes with
the electronic levels present in the dot itself. Such hybridization
would also be harmful to the prospects of constructing the SYK model.  Here the authors
show that a remarkable instance of  level repulsion takes place and
protects the Majorana modes. Random matrix theory considerations imply
that a
generic model describing the situation depicted in Fig.\ \ref{fig6} has the
the spectrum composed of the zero-mode manifold ($N$ states)
separated by $\epsilon\approx N\delta\epsilon_{\rm typ}/\pi$ from all
other energy levels. Here $\delta\epsilon_{\rm typ}$ is the typical
energy level spacing in the dot before it has been coupled to the wires. 
 
The key advantage of this proposal is that Majorana zero modes are now
routinely observed in proximitized semiconductor quantum
wires. Assembling a large number of such wires to form a device in
Fig.\ \ref{fig6} is clearly a significant challenge.  However, recent history of
this field  has proven that even tough chalenges can be met when the
goal is deemed sufficiently important. 

\subsection{SYK model at the surface of a 3D topological insulator}
Another canonical platform for Majorana zero modes is the so-called
Fu-Kane superconductor, which arises from proximitizing
the topologically protected surface state of a 3D strong topological
insulator (TI). Magnetic vortices in the Fu-Kane superconductor have been
predicted \cite{FuKane2008} to harbor Majorana zero modes. Signatures
consistent with this prediction have recently been reported in
Bi$_2$Te$_3$/NbSe$_2$ heterostructures \cite{Jia2016a,Jia2016b}.

\begin{figure}[t]
\includegraphics[width = 7.6cm]{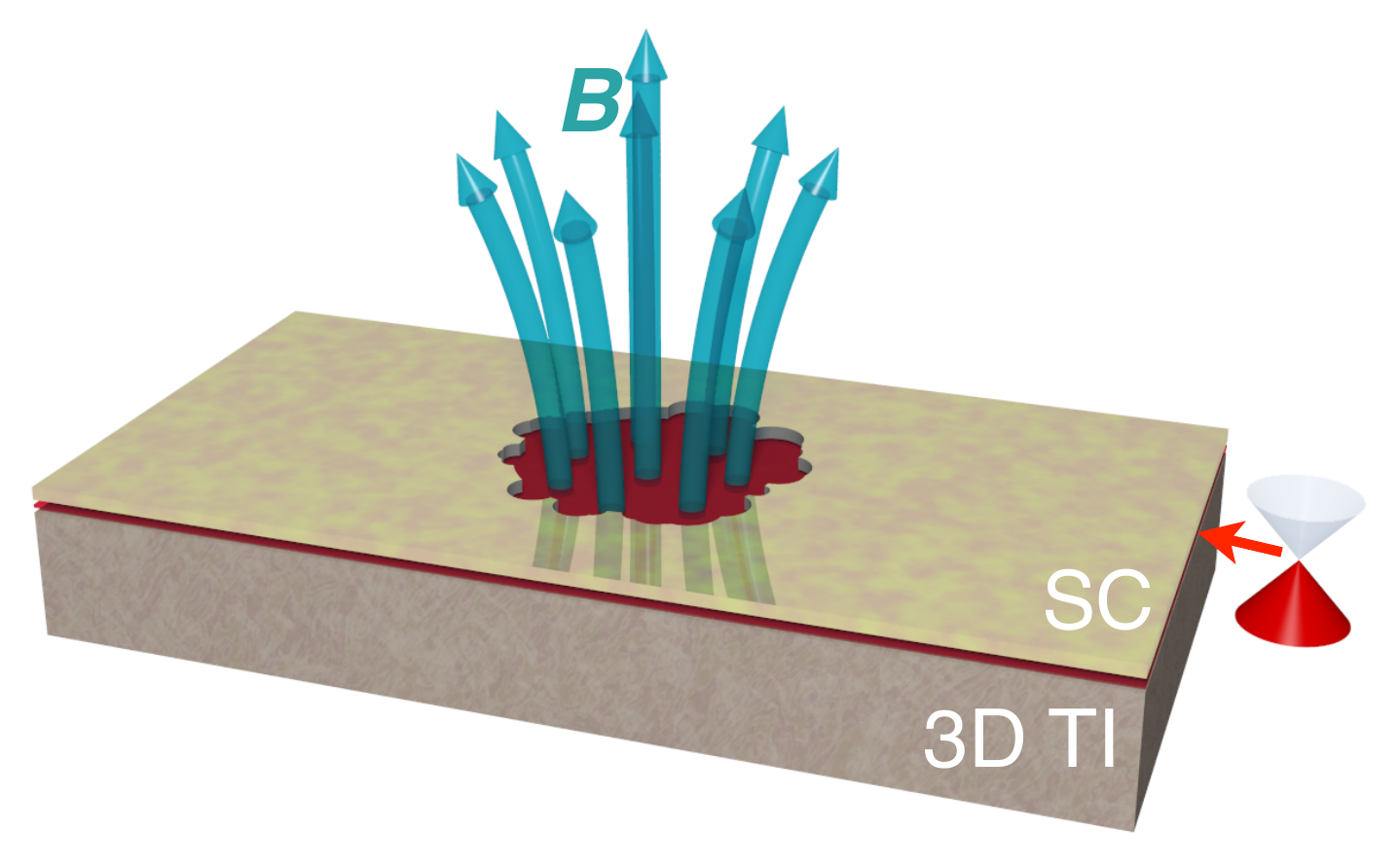}
\caption{{\bf Schematic of the SYK model realization in the Fu-Kane superconductor.}
A 3D topological insulator surface has been proximitized by depositing
on it  a thin film of a superconducting material. An irregular-shaped hole in the
film serves to trap magnetic flux. For each magnetic
flux quantum trapped there is one Majorana zero mode localized in the hole.  
}\label{fig7}
\end{figure}
Realization of the SYK model in this platform has been proposed in
Ref.\ \cite{Pikulin2017}. As with the quantum wires the key challenge in
this setting is to assemble $N$ Majorana modes in the same spatial
region without producing bilinear hybridization terms. This is
achieved by engineering a nanoscale hole in the superconducting film
on the surface of a TI. The hole serves to trap multiple vortices,
each binding a Majorana zero mode. The mechanism behind this trapping has to
do with the well known phenomenon of vortex pinning: since the
superconducting order parameter $\Delta$ is suppressed  to zero at the core of
a vortex it is energetically preferable to position the core in a
region where $\Delta$ was already suppressed by other effects
(impurities for example). A hole in the superconductor has $\Delta=0$
and thus serves as a perfect
pinning center. When sufficiently large it can trap many vortices, and
thus many Majorana zero modes.

Hybridization between the zero modes can be avoided in this setup by tuning the
chemical potential $\mu$ of the TI to the so called neutrality point. The
topologically protected  surface state in a TI is described as a
single massless Dirac fermion in 2D, familiar from the physics of
graphene \cite{Neto_RMP}. At neutrality $\mu$ coincides with the Dirac
point. Importantly the Fu-Kane superconductor acquires an extra
symmetry at this special point \cite{TeoKane2010} that acts exactly
like ${\tilde\cT}$ defined in Eq.\ (\ref{dot1}). This symmetry then
excludes any bilinear terms from the effective Hamiltonian describing
the zero modes. If interactions are present between the underlying
electron degrees of freedom then the leading term in such an effective
description is a 4-fermion SYK-like Hamiltonian (\ref{hsyk}). Using a
combination of analytical and numerical tools Ref.\ \cite{Pikulin2017}
 showed that when the hole is engineered to have a highly irregular
 shape Majorana wavefunctions acquire random spatial structure and the
 screened Coulomb interaction between electrons produces essentially
 random couplings $J_{ijkl}$, as required to implement the SYK model.

In comparison to the quantum wires experimental understanding of the Majorana
modes in the Fu-Kane superconductor remains significantly less
developed. Results reported in \cite{Jia2016a,Jia2016b} have yet to be
reproduced by another group or in another family of materials. This
makes it more difficult to ascertain the prospects for realization of
the above proposal. On the other hand once the Fu-Kane superconductor has
been better understood and characterized, it should be relatively easy
to fabricate and probe the  device proposed in Ref.\ \cite{Pikulin2017}. Also, the fact that
bilinear terms can be controlled by tuning a single parameter (the
chemical potential $\mu$) confers some advantage over the quantum wire
realization, where the smallness of such terms relies on an
approximate symmetry of the system that cannot be easily manipulated.

\section{Conclusions and outlook}

In this review we focussed on proposals for experimental realizations of the cSYK or  SYK model. 
All such proposals currently on the table face significant challenges. These include
hurdles in materials synthesis, device fabrication, reproducibility
and control.  Nevertheless the field evolves rapidly and this provides
hope that some version of the theoretically proposed devices could be experimentally
realized in the near future.

Several measurements designed to make the connection to quantum black holes explicit have been proposed, and they present their own challenges. Measurement of out-of-time-ordered quantities,
important for the identification of many-body chaotic behavior
characteristic of a black hole, constitutes a huge challenge in the
atomic physics setup and remains an unsolved problem for all proposed solid
state realizations. Spectroscopic or transport identification of the
emergent  black hole physics is more straightforward in principle but
by no means free of challenge.

Despite these technical hurdles, the key conceptual advances achieved over the past two years outline a clear roadmap
which, if successfully followed, should lead to experimental insights to some of
the most fundamental mysteries facing modern physics.  
The realization that a simple, solvable quantum mechanical model
exhibits fundamental connections to quantum black holes has brought quantum gravity into the
realm of being potentially experimentally testable. This exciting prospect advances the decades-long quest, started already by Einstein, to
reconcile two fundamental theories of nature, general relativity and
quantum mechanics. That deep ideas on the nature of quantum gravity could be ultimately tested in a
humble piece of solid replete with disorder and imperfections reinforces the modern belief in the principle of emergence, that
indeed ``more is different'' \cite{Anderson1972}.

Speculatively, but most importantly, once an experiment has produced a
system which can be reliably identified as a quantum black hole, one
can turn to empirically investigating many subtle questions
pertaining to quantum gravity.  The effort to formulate such
longstanding questions in the context of simple quantum mechanical
model is a recent and fast developing subject, see for example
\cite{Cotler:2016fpe,Maldacena:2017axo,
  Kourkoulou:2017zaj,Maldacena:2017axo, Maldacena:2018lmt} for some
examples of such attempts to address quantum gravitational questions
using the SYK family of models.

\begin{acknowledgments}

The authors are indebted to many colleagues who helped shaped their
understanding of the subject. Of these special thanks go to
I. Affleck, E. Altman, J. Alicea, L. Balents, M. Berkooz, A. Chen, F. Haehl, A. Kitaev,
C. Li, E. Lantagne-Hurtubise, P. Narayan, D. Pikulin, S. Sachdev, J. Simon and M. Tezuka.
\end{acknowledgments}

\bibliography{SYK}

\end{document}